\documentclass[preprint,aps,prl,showpacs,amsmath,amssymb]{revtex4}
\usepackage{hyperref}

\usepackage{epsfig}
\usepackage{graphicx}
\usepackage{dcolumn}
\usepackage{bm}

\begin{document}
\title{A dual modeling of political opinion networks}
\author{R. Wang$^{1,2}$ \footnote{ E-mail address: wangr@ismans.fr} and
  Q.A.Wang $^{1,3}$ \footnote{E-mail address: awang@ismans.fr}}
\affiliation{$^1$LUNAM Universit\'e, ISMANS, Laboratoire de Physique Statistique et Syst\'emes Complexes, 44, Avenue, F.A. Bartholdi, 72000, Le Mans, France. \\
$^2$College of Information Science and Engineering, Huaqiao University, Quanzhou, 362021, P.R.China\\
$^3$LPEC, UMR CNRS 6087, Universit\'e du Maine, 72085 Le Mans, France}

\date{\today}

\begin{abstract}
We present the result of a dual modeling of opinion network. The model complements the agent-based opinion models by attaching to the social agent (voters) network a political opinion (party) network having its own intrinsic mechanisms of evolution. These two sub-networks form a global network which can be either isolated from or dependent on the external influence. Basically, the evolution of the agent network includes link adding and deleting, the opinion changes influenced by social validation, the political climate, the attractivity of the parties and the interaction between them. The opinion network is initially composed of numerous nodes representing opinions or parties which are located on a one dimensional axis according to their political positions. The mechanism of evolution includes union, splitting, change of position and of attractivity, taken into account the pairwise node interaction decaying with node distance in power law. The global evolution ends in a stable distribution of the social agents over a quasi-stable and fluctuating stationary number of remaining parties. Empirical study on the lifetime distribution of numerous parties and vote results is carried out to verify numerical results.

{\it keywords}: Complex network, opinion network, dual model, network of political parties

\end{abstract}

\maketitle

\section{I. INTRODUCTION}

The network approach has contributed significantly to our understanding of the structure and properties of various complex systems ranging from transport networks to human society\cite{s1}-\cite{s4}. In view of the complexity of the dynamics in these systems, considerable simplification is necessary in the numerical and analytical modeling. A typical example of the approach is the agent-based modeling of (political or economical) opinion networks\cite{s4}. The process of opinion formation has been modeled as a collective dynamics in which the agents update their opinions by local majority or by imitation\cite{s5}. The agents are located on the nodes of a graph and endowed with a finite number of discrete or continuously variable opinions (e.g.,  two opinions with spin up and spin down)\cite{s6}-\cite{s8}. Both numerical and analytical results showed that the system could reach a total consensus or a state of coexistence of different opinions. In the case of continuous opinions, the system can reach a state of polarization or fragmentation with a finite number of groups having different opinions\cite{s9}-\cite{s10}.

Interest has grown recently in a new type of network named adaptive network\cite{s11}. They are characterized by the existence of the feedback loop between the opinion dynamics on the network and the structural dynamics of the network, meaning that the network evolves in time by rewiring the links due to the opinion change of the node, and by modifying opinion due to the topological structure evolution\cite{s12,s13,s14}. It was observed in the numerical results that, in the case of continuous opinions, there was cluster formation by enhancing communication between agents of similar opinions, while the adaptability of the topology structure favors the division of these clusters\cite{s14}.

The results obtained with these models are useful for understanding certain universal,  global and statistical features of opinion systems. A possible improvement is to make the models closer to the reality by considering additional dynamical variables involved. For example, the different opinions are not only the attribute of the social agents, but also the elements of a network apart in which an opinion dynamics (birth, death, union and division etc.) can occur and influence the opinion adhesion of the social agents. Similarly, the social agents composing the voter network have also its own dynamics (rewiring, breaking, and addition of links, change of opinion etc) which certainly affects the dynamics of the opinion subnetwork. The social agents are in addition elements of other networks (cultural, economic, professional, environmental, ecological and so forth) and constantly adapt their behavior and features to these "environments". All these subnetworks form the global network of the society which is self-adaptive or self-referential through separate dynamics of the different self- or mutual-adaptive subnetworks.

Based on this idea, we construct a dual model containing two interacting networks, one being composed of a large number of social agents, another one a small number of political parties or social opinions. The influence of the other social entities and networks such as economy, culture, international situations whose effect is in general a random factor for long term point of view, is simplified as a mean field or the "temperature" of a thermostat which can be controlled or changed randomly. This model is illustrated in figure 1. Certainly, this is a simplified version of a small part of the society, but it consists in a step forward in the description of the real network of political, economical or cultural opinions.

\begin{figure}[htb]
\begin{center}
\includegraphics[width=0.65\textwidth]{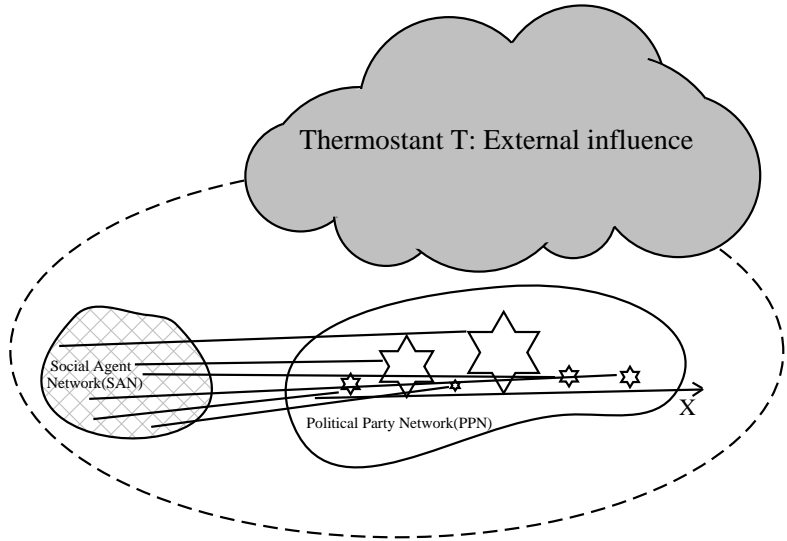}
\caption{An illustration of the dual model containing two interacting subnetworks with the external influence represented by a thermostat. One subnetwork is composed of a large number of social agents while another of a small number of political parties (stars) initially distributed randomly on the x-axis in the range $\left[0,1\right]$ representing the political spectrum. }
\label{fig1}
\end{center}
\end{figure}

\section{II. DESCRIPTION OF THE MODEL}
This dual model includes two subnetworks. The first one is the network of opinions or political parties. It initially contains a number $n$ of political parties randomly distributed on a one dimensional axis with discrete opinion position $x_{i}$ ($i=1,2...n$) (political spectrum) in the interval from $0$ to $n$. Any two political parties, $i$ and $j$, can have interaction or correlation such as communication, mutual support, exchange of viewpoints and members, mutual adaptation and so on. These interactions are simply modeled by a strength $S_{c}(ij)$ given by:
\begin{equation} \label{eq1}
S_{c}(ij) = \frac{1}{|(x_{i}-x_{j})/n|^{\alpha}}-1,
\end{equation}
\noindent where $\alpha>0$ is a parameter characterizing the interaction property. The distance $|x_{i}-x_{j}|$ will be divided by $n$ (total number of parties or the length of $x$ axis) to make the strength independent of the network size. The magnitude of the strength is evaluated with respect to the extreme left and the extreme right who do not have interaction due to the term $-1$. This is of course a simplified version of the reality (to quantify the interaction strength) since the extremists actually do have some relationships depending on their distance.

The interaction may be affected by the social environment. Since this environment is "averaged" with the temperature $T$ of the thermostat, we model its effect on each party $i$ in an average way with respect to the opinion mean $\bar{x}$\cite{Comment1} through an exponential factor in the interaction strength:
\begin{equation} \label{eq2}
S_{c}^{T}(ij) =\left(\frac{1}{|(x_{i}-x_{j})/n|^{\alpha}}-1\right)e^{-x_{i}( \frac{1}{T}-\frac{1}{\bar{x}})}.
\end{equation}

\noindent This is an imitation of the temperature effect in the Boltzmann distribution of equilibrium statistical mechanics, so that increasing temperature will enhance the interaction between this party and a given party $j$. Clearly, if $T=\bar{x}$ at an "equilibrium" between the system and its environment, no external influence occurs. If $T>\bar{x}$ (or $<\bar{x}$), the interaction is more strengthened for larger $x_i$ (or smaller $x_i$). Put it differently, higher (or lower) temperature favors the interaction to the right (or left). If the political spectrum was symmetrical, e.g., from $-0.5$ to $0.5$, the above effect to the right and left would be symmetrical. But in this work, taking into account the fact that the left and right wings are not symmetrical in general from political viewpoint, we keep the spectrum positive, i.e., from 0 to 1, so that $T>\bar{x}$ and $T<\bar{x}$ produce different effects on the left and right.

One of the essential properties of an opinion or political party is its attractivity. This property depends on many features such as viewpoint, political strategy or tactics, communication skills, eloquence of the representatives or speakers etc. This attractive character is represented in our model by the eloquence $E_{i}$ of a party $i$ whose values in the interval [1,10] are initially distributed randomly to the parties.

The age of an opinion or party is also an important phenomenological factor for its evolution. In this model, each party increases its age by one at each step of the monte-carlo simulation. If two parties merge into one, both are considered dead with a newborn at the same time. If a party splits into two parties, it is considered dead accompanied by two newborns.

The dynamics in the opinion subnetwork includes merging and splitting in the following way.

\begin{enumerate}
\item Two parties, $i$ and $j$, can merge into a new one with a probability proportional to the interaction strength $S_{c}(ij)$ and to the sum of the numbers of their supporters $N_{i}+N_{j}$, i.e., $p_{m}\propto S_{c}(ij)*(N_{i}+N_{j})$. The position of the new party is located at $x_{l}=\frac{x_{i}+x_{j}}{2}$\cite{Comment2}. The eloquence of this new party takes the larger value of the two merged parties. The age $A_{l}$ of the newborn is 1.

\item A composite party $l$ (composed of merged parties) can split into two parties $i$ and $j$ with a probability given by $p_{s}\propto N_{l}*A_{l}$, meaning that more aged party with larger number of supporters is more possible to split. While one of the party $i$ stays in the original position $x_i=x_{l}$ and keep the majority of the original agent number, the another one $j$ takes a new position $x_j$. The distance $\left|x_i-x_{j}\right|$ is determined with a positioning probability proportional to the interaction strength $S_{c}(lj)$ (or $S_{c}^{T}(lj)$ with the thermostat effect). Concerning the number $N_{j}$ of agents belonging to this new party, we suppose that it takes $5\%$ of the agents belonging to each party $g$ with a probability $p(jg)\propto S_c(jg)E_jN_g$, i.e., $N_{j}=\sum_g N_{g}p(gj)*5\%$, where the eloquence of the newborn $E_j$ randomly takes a value between 1 and 10. The two new born parties have the age of 1.

\end{enumerate}

The second subnetwork containing $N$ social agents is initially a random graph with an average degree $\bar{k}=N*p_{c}$ (Erdos-Renyi network)\cite{Newman}. This subnetwork is in correlation with the subnetwork of parties through the opinion adhesion of the agents. At $t=0$, this adhesion is given by either a Gaussian or a uniform distribution of agents over the political spectrum.

At each time step $t$, with equal probability, the agent network evolves with three dynamics at the same time:

\begin{enumerate}
\item An agent belonging to a party $i$ is randomly chosen and linked to another agent of a different party $j$ with a probability proportional to the interaction strength $S_{c}(ij)$ between the two parties and to the number of agents $N_{j}$ belonging to the party $j$, i.e. $p_{a}\propto N_{j}*S_{c}(ij)$. The first mechanism implies that the interaction between two parties can favor the contact between their supporters. The second one is a kind of preferential adhesion depending on the number of agent of the considered party.

\item The link between a pair of randomly chosen agents belonging to different parties $i$ and $j$ can be broken with a probability inversely proportional to the interaction strength between their respective parties $p_{b}\propto 1/S_{c}(ij)$, meaning that the interaction between two parties can contribute to maintain the contact between their supporters.

\item A randomly chosen agent can change his opinion (political adhesion) from the party $i$ to the party $j$ with probability given by $p_{r}\propto S_{c}(ij)*E_{j}*N_{j}$, where $E_j$ is the eloquence of the party $j$. This construction of adaptive agent subnetwork is based on the idea that each voter has the possibility to change his opinion under the influence of the opinion attractivity and of what the other agents are doing (social validation\cite{Sznajd}).

\end{enumerate}

When the effect of the thermostat is taken into account, the interaction strength should be replace by $S_{c}^{T}(gh)$ in the above dynamics.

In general, the dynamics in the two subnetworks has different time rate or time scale. We use the time ratio $TR$ to characterize the temporal difference, i.e., one time step in the dynamics of opinion network corresponds to $TR$ time steps in the social agent dynamics.

\section{III. RESULT AND DISCUSSION}

\subsection{A. Distribution of Agents over Parties}
In the simulation, the size $N$ of the agent subnetwork is a constant. The isolated agents and duplicate links are not allowed, and each agent can only adhere to one party at any time. The political spectrum is normalized (devided by $n$) so that it is independent of the size of the opinion subnetwork. We used two initial distributions of agents over political spectrum: the normal and the uniform distribution (Figure 2), for the same setting of parameters $n=500,N=5000,TR=2,\alpha=0.1$.

It is observed (Figure 2) that the system evolves to a stable distribution of agents over parties with some peaks in the middle range of political spectrum, and this is independent of the initial distribution. This final normal-like distribution of agents can be explained by the dynamical rules. The interaction tends to unify the parties whose positions are close and to create new one situated between them. Hence the parties in the middle range of the political spectrum have more chance to merge and to split with the newborns situated around them. As shown below, the number of the remaining parties decreases in the evolution due to the merging dynamics. As a consequence, the number of agents of each party increases mainly in the middle range of the spectrum.

In order to see if this final state is in accordance with the reality, we carried out an empirical study of the distribution of voters over the different parties in the parliamentary elections mainly in the European countries. The information about the election ballots and the political spectrum of the party are collected from the website \cite{s15}. The empirical result shown in Figure 3 in comparison with the numerical result is averaged over 11 countries, 159 parties and 186 elections. The data in Figure 3(b) shows that most voters are located around the middle of the political spectrum and few voters for the positions approaching the two extreme points, which statistically confirms the numerical result of the model. Notice that this result is obtained without considering the influence of the thermostat.

\subsection{B. The Opinion Subnetwork}
Let us now turn to the subnetwork of parties. As shown in Figure 4, the number of parties first decreases and then reaches a quasi-stable state with fluctuation. The inset of Figure 4 shows straight lines in log-linear scale, meaning that the evolution of the number of parties approximately has an exponential form. The relaxation time before reaching the quasi-stable state depends on both the time ratio ($TR$) and the initial number of political parties. Larger $TR$ and initial number of parties result in longer relaxation time. This is reasonable since large $TR$ slows down the evolution of the opinion subnetwork with respect to the agent subnetwork whose evolution decides the time course of the whole network, and large number of parties need more time to be reduced to the quasi-stable numbers.

The reason for the fluctuation of the number of remaining parties in the quasi-stable state is that the system periodically goes through splitting and merging. In average, the fluctuation period increases with increasing $TR$ as shown in Figure 5.

It is observed that the time average number $<n_t>$ of remaining parties depends on the exponent $\alpha$ of the interaction strength and on the initial number of parties, where $n_t$ is the number of parties at time $t$. This is shown in Figure 6. Larger initial number of parties results in larger final number of parties for fixed size of agent subnetwork ($N=5000$), and larger $\alpha$ leads to smaller final number of parties. This later effect is certainly the consequence of the party union enhanced by the interaction strength which is larger for larger $\alpha$ at the same distance since the normalized distance is smaller than 1. In Figure 6(b), one observes a sharp drop of $<n_t>$ from around 20 to 2 at about $\alpha=0.3$.

The lifetime of the defunct parties is recorded and compared with the empirical data we collected from the website \cite{s16} with $645$ defunct parties. In order to compare the time scales of numerical and empirical data, we normalized the lifetime by dividing it with the maximum lifetime of each data. It is found (Figure 7) that the numbers of the defunct parties are distributed over lifetime in a power-law with an exponent close to 2, which is similar to the heavy tail of waiting time distribution in human dynamics\cite{Nature}.

\subsection{C. The structure of Agent Subnetwork}

In order to study the evolution of the topological structure of the agent subnetwork under the influence of the merging/splitting dynamics of the party subnetwork and of the link adding/deleting and opinion change dynamics of the agents, we studied a quantity called solidarity $S_{i}$ of a party $i$ which is defined for each party as the ratio between the number of links among the agents of this party to the number of links these agents have with the agents of other parties. It is observed that there are three typical steps in the time evolution of $S_{i}$ indicating a gradual formation of political community structure and the variation of the community character with the dynamics of the opinion subnetwork. In order to show this, the time evolution of the maximum solidarity $(S_{i})_{max}$ and average solidarity $<S_{i}>$ is drawn in Fig.8.

At $t=0$, the maximum and average solidarity are in the order of $10^{-3}$. Then there is a quick increase of $(S_{i})_{max}$ up to a climax of about 3.2 around $t=3200$ followed by a steep descent down to about 0.2. This culmination (more or less sharp) happens later on approximately every 200 steps which corresponds to the time period of the fluctuation of party numbers in the quasi-stable state. The sharp peaks correspond to the smallest numbers of parties. In other words, the periods of increasing (decreasing) $(S_{i})_{max}$ correspond to periods of merging (splitting) of the parties. The behavior of $<S_{i}>$ is similar to that of the $(S_{i})_{max}$. This similarity can be explained as follows. Since most of the parties have very small $S_{i}$, and only few parties have very large $(S_{i})_{max}$, the average solidarity is mainly determined by the solidarity of the large parties.

Since the initial values of $S_{i}$ are all about $10^{-3}$ in the random agent subnetwork (no political influence), the increasing $S_{i}$ up to values much larger than the initial ones means a politicization of the agent subnetwork in the evolution, or a formation of political community. This politicization is particularly strong in the periods when there are few parties.

It is also found that the average degree $\bar{k}$ of the agent subnetwork is almost a constant during the evolution. The relative average degree $ \bar{k_{i}}/\bar{k}$ for each party, i.e., the average degree of the agents belonging to the party $i$ divided by the average degree of the whole agent subnetwork, is always close to 1. Figure 9 shows the time evolution of the maximum relative average degree $(\bar{k_{i}}/ \bar{k})_{max}$ and of the mean of relative average degree $\overline{\bar{k_{i}}/ \bar{k}}$, both of them decrease first and then fluctuate around 1. This result indicates that, in average, the total number of links in the system does not change despite the dynamics of link adding/deleting and the formation of political community. In average, each agent has more or less the same degree in the evolution.

The above result implies that there is a redistribution of links in the evolution. Initially, the links is uniformly distributed among all agents at any distance and, later on, are redistributed mainly among the agents belonging to the same parties. This aspect of the community formation can be seen in the evolution of a quantity we called political distance between the linked agents which is defined for a pair of linked agents by the distance between their respective parties. Figure 10 shows the political distance distribution of links at three typical time steps. The distribution changes from the initial semi-Gaussian distribution to the final semi-delta type distribution around zero distance, a strong sign of the politicization of the population or formation of political community. This result is reasonable if one considers the dynamics of adding/deleting links regulated by the interaction strength between the parties, i.e., the evolution favors the links between agents having the same opinions.

\subsection{D. Effect of Thermostat}

In addition to the interaction between the agent subnetwork and opinion subnetwork, the influence of the evolution of other social networks plays an important role in opinion formation. This environmental influence is a kind of external force represented in this dual modeling by the temperature of a thermostat which enters into the interaction strength through the exponential factor $f(T)=e^{-x_{i}(\frac{1}{T}-\frac{1}{\bar{x}})}$ in Eq.(\ref{eq2}). When the ``temperature'' $T$ of the thermostat is larger (or smaller) than average opinion, this factor is larger (or smaller) than 1, which yields a larger (or smaller) interaction strength for larger $x_i$ position.

Figure 11(a) shows the evolution of average opinion with two different temperatures in comparison with the result without thermostat. The average opinion (initially at 0.5) shifts to the left ($\bar{x}<0.5$) when $T<0.5$ and to the right ($\bar{x}>0.5$) when $T>0.5$. However, it always ends around 0.4 without thermostat. The maximum average opinion is $x_{max}=0.6$ no matter how high $T$ is. However, at low temperature, the minimum average opinion can be very close to the extreme left.

The thermostat also has influence on the final numbers of parties. As the Figure 11(b) shows, higher $T$ leads to larger final number of remaining parties. As $T$ is larger than $0.2$, the average number of remaining party is $<n_t>\approx 26$ with $\alpha=0.1$, $N=5000$, $n=500$, and $TR=2$. $<n_t>$ is 2 when $T$ is below 0.2. This result is in accordance with the result in Figure 11(a), lower temperature $T$ leads to a small number of remaining parties located on the left side of the political spectrum. When the temperature is above 0.2, in average more than 20 remaining parties are located mainly on the right side of the political spectrum. The jump of $<n_t>$ around $T=0.2$ happens when $T=\bar{x}$. This explanation is in agreement with the effect of the thermostat on the interaction strength mentioned above, i.e., $T>\bar{x}$ favors the shift of the parties and agents to the right, while $T<\bar{x}$ favors their shift to the left. In other words, this "transition of political regime" from a "strong democracy" (more than 20 remaining parties in average) to a "weak democracy" even a "dictatorship" (2 remaining parties in average) is a consequence of the exponential factor of the thermostat of this model.

The small number of parties for $T<\bar{x}$ can be explained by the large merging probability due to the small distances between the parties squeezed by the low temperature in a small region close to the the extreme left. The high temperature does not have the same squeezing effect to the right due to the asymmetrical property of temperature mentioned in section II.

The thermostat does not change sensibly other aspects such as the relaxation time, lifetime distribution and the structure of agent network.

\section{IV.  CONCLUDING REMARKS}

In this work, we have proposed a dual modeling of opinion network. The model is composed of two interacting subnetworks, each having its own evolution mechanisms. In the opinion subnetwork, two parties can merge into a new one, and any composite party can split into two parties. In the agent subnetwork, opinion change and link adding/deleting take place. The above dynamics occurs in a probabilistic way depending on the interaction strength between the political parties, their age, and their attractivity. The global network is in contact with a thermostat representing the environmental influence on this dual network.

The numerical results of the model for the opinion subnetwork can be summarized as follows. When the thermostat is not taken into account, the party number decays exponentially from initial number to a quasi-stable regime in which the party number fluctuates quasi-periodically with alternative merging and splitting periods. The average number of remaining parties is around 20 for initial number $n=500$ and the exponent $\alpha<0.3$. This average number is 2 for $\alpha>0.3$. The lifetime distribution of the parties given by this model was verified with empirical data.

If we call the state of 20 parties a strong democracy regime and the state of 2 or 1 parties a weak democratic or dictatorship regime, there is a transition between these two regimes at about $\alpha=0.3$ driven by an order parameter $\alpha$ which is the exponent in the interaction strength between parties. This same transition between two political regimes also takes place with temperature of the thermostat when this latter is considered and $\alpha<0.3$. The critical temperature is the average opinion $T_c=\bar{x}$. Another effect of the thermostat is that very low temperature squeezes a very small number of parties (weak democratic regime or dictatorship) to the extreme left region, and very high temperature pushes numerous remaining parties to be dispersed in the whole right wings.

On the other hand, for the agent subnetwork, when the thermostat is not considered, the system ends in a stable distribution of agents over the political spectrum with majority of the agents in the middle of the spectrum, no matter what is the initial condition. This is mainly the result of the interaction strength $S_{c}(ij)$ which favors the merging dynamics of the parties close to each in the middle of the political spectrum and the concentration of agents in that region. This stable distribution is in accordance with empirical data.

Another important feature of this model is the time evolution of the politicization of the agents from the initial non political topological structure of the population. This political community formation and its time fluctuating character are consequences of the merging/splitting dynamics. The politicization is extremely strong in the periods or regime where there are few remaining parties. The quantities "solidarity" and "political distance" are used to characterize the degree of this politicization.

This model is an effort to make more realistic modeling of opinion networks on the basis of the previous agent-based opinion network. Although some general features are new, such as the regime transition, the party's lifetime distribution, solidarity, external influence and so forth, many features are similar to the consensus dynamics of agent-based models. For example, the strong (weak) politicization can be seen as a strong (weak or poly-) consensus state. Many general features and characteristics of the dynamics are in accordance with empirical data or phenomena of the society. Examples include the party lifetime distribution, voter distribution over parties and the exponential relaxation. The politicization of the population with quasi-periodical oscillation is also an general phenomena of our society. All these results cannot be purely coincidence and should be the consequence of some realistic dynamical mechanisms included in the model. The present work used only a very basic interacting dual network modeling. More realistic improvements are feasible in this approach by further consideration of, for example, more variable in the political spectra as proposed\cite{Sznajd}, more subnetworks, multiple thermostats representing different influences from culture, economy, natural disasters, international environment etc, and more general dependence of the involved probabilities on the various parameters.

\textbf{Acknowledgments}:
We thank Cyril PUJOS for the contributions to the simulation programs.  This work is supported by the Region des Pays de la Loire of France under the grant number 2009-9397 and and Research Foundations of HuaQiao University ( No. 09BS511).






\begin{figure}[htb]
\begin{center}
\includegraphics[width=0.9\textwidth]{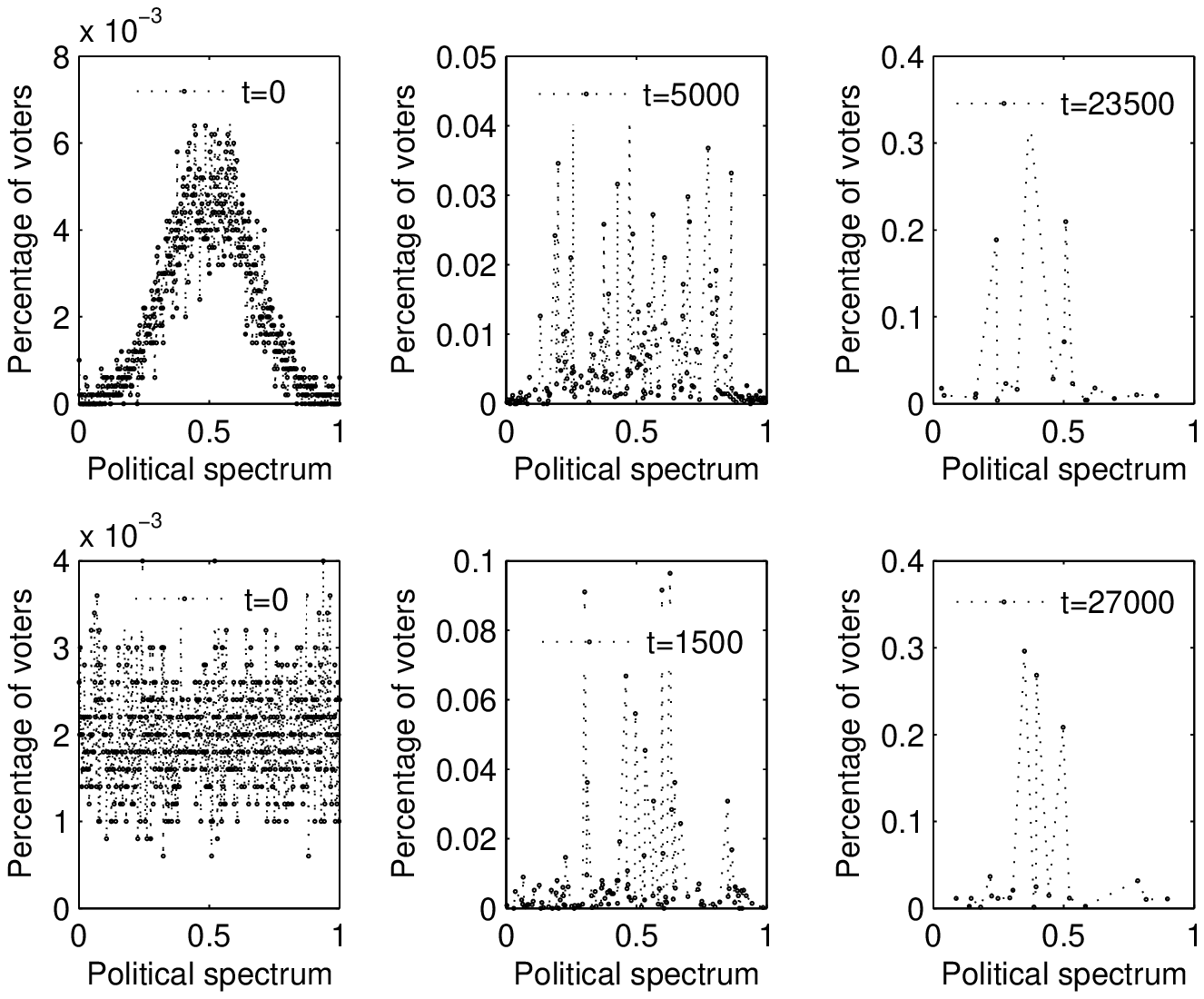}
\caption{The evolution of agent distribution over political parties for the initial normal distribution (the top three figures) and the initial uniform distribution (the bottom three figures). The final stable distribution with most of the agents in the middle range of the political spectrum is independent of the initial distributions. }
\label{fig1}
\end{center}
\end{figure}

\begin{figure}
\begin{center}
\includegraphics[width=0.75\textwidth]{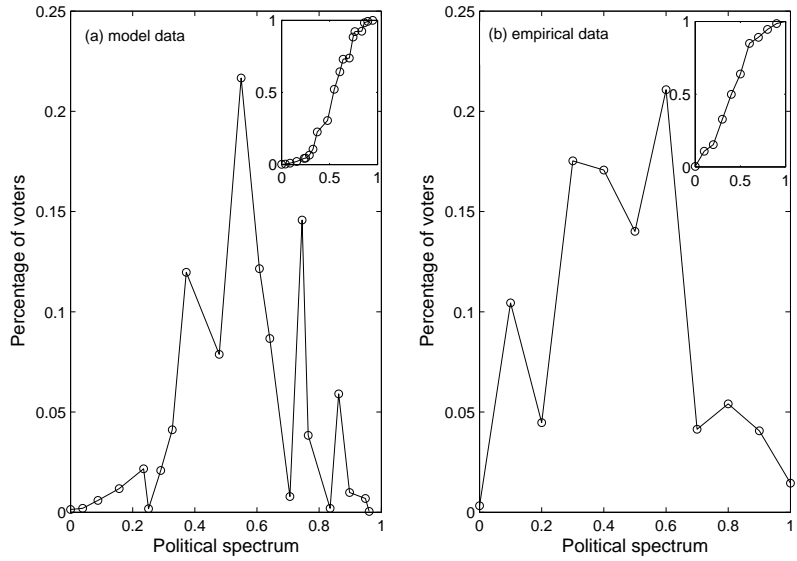}
\caption{Distribution of agents (voters) over the political spectrum given by (a) the dual model with parameters $n=500$, $N=5000$, $TS=2$, $\alpha=0.1$ and (b) empirical data collected from 11 countries, 159 paries and 186 elections. The insets are the corresponding cumulative distributions. }
\end{center}
\end{figure}

\begin{figure}
\begin{center}
\includegraphics[width=0.75\textwidth]{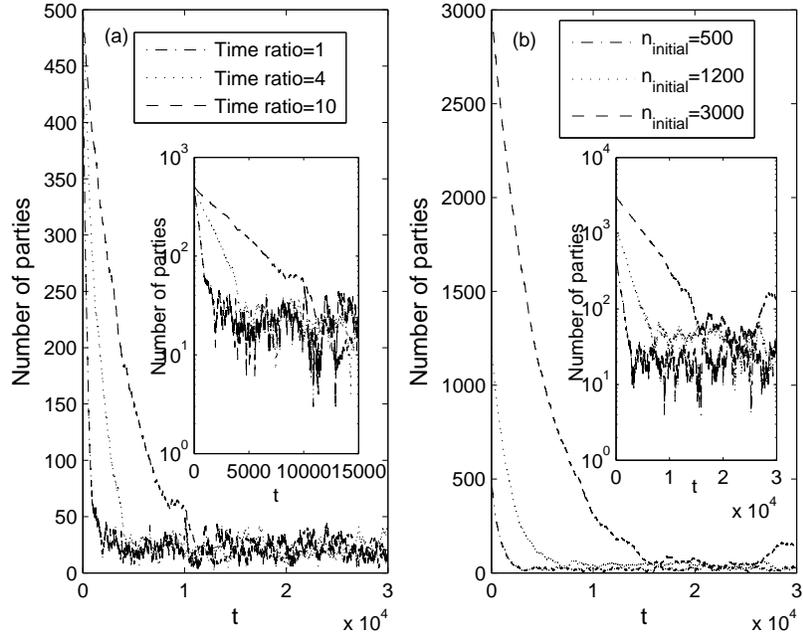}
\caption{The evolution of the number of political parties for (a) different time ratio with $N=5000, n=500, \alpha=0.1$ and (b) different initial number of parties with $N=5000, \alpha=0.1, TR=2$. The nearly straight lines in the insets in log-linear scale indicate an exponential decay of party number.}
\end{center}
\end{figure}

\begin{figure}
\begin{center}
\includegraphics[width=0.75\textwidth]{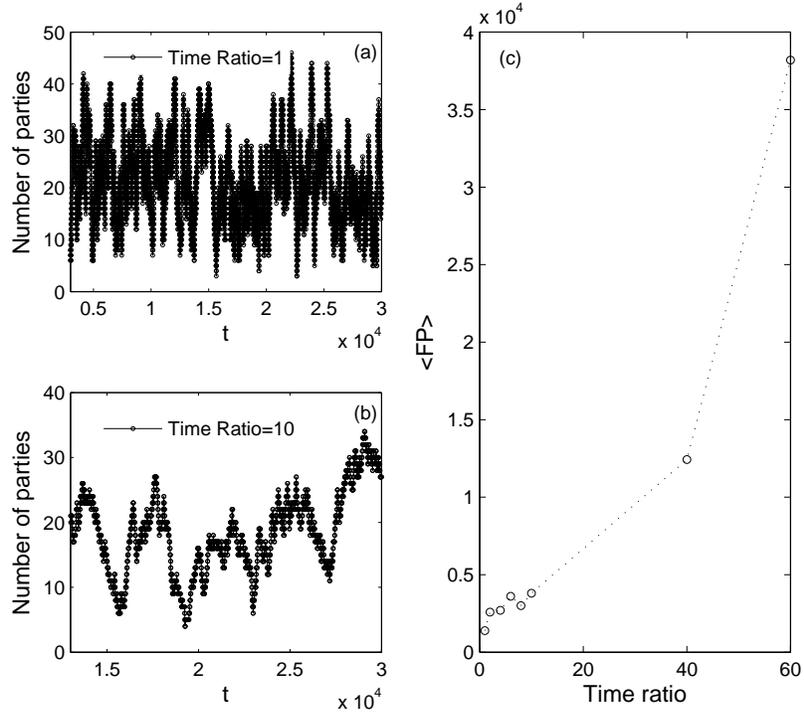}
\caption{The fluctuation of number of parties when the system reaches the quasi-stable state with different time ratio (a) $TR=1$ and (b) $TR=10$. The figure (c) shows the increase of the average fluctuation period $<FP>$ with increasing time ratio. }
\end{center}
\end{figure}
\begin{figure}
\begin{center}
\includegraphics[width=0.75\textwidth]{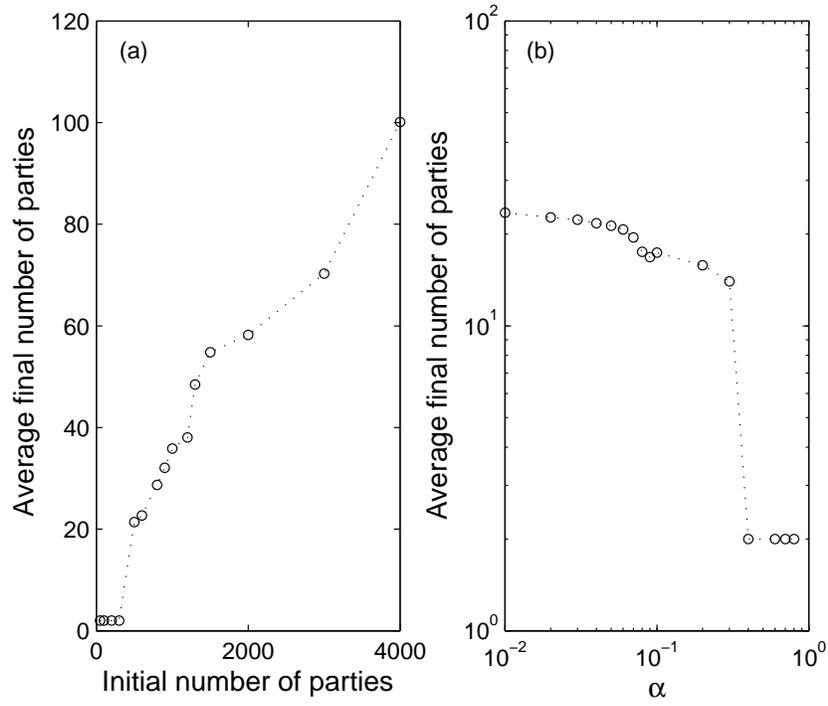}
\caption{The average number $<n_t>$ of parties at the quasi-stable state as a function of (a) the initial number of parties and (b) the exponent $\alpha$ of interaction strength. The sharp drop in $<n_t>$ in (b) occurs at about $\alpha=0.3$.}
\end{center}
\end{figure}

\begin{figure}
\begin{center}
\includegraphics[width=0.75\textwidth]{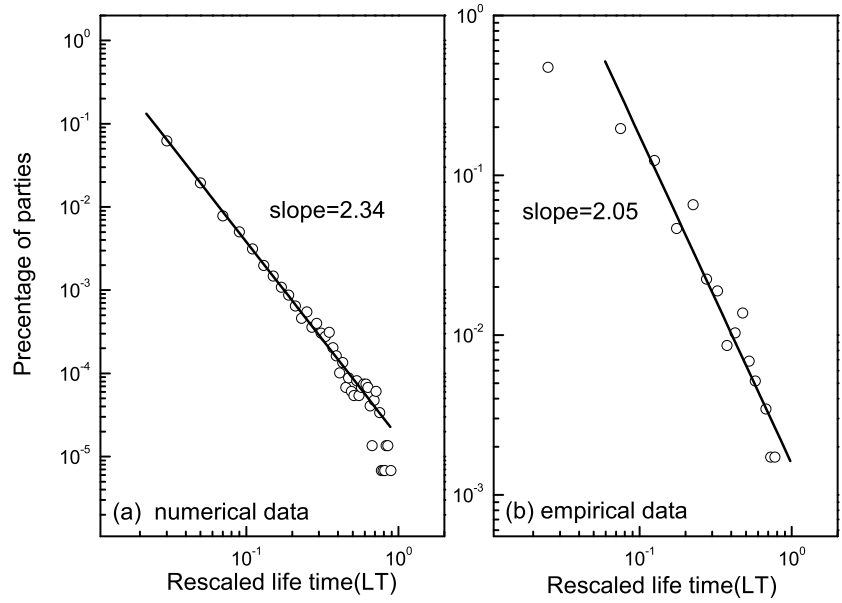}
\caption{The lifetime distribution of parties: (a) numerical result of the dual model and (b) empirical data collected with 645 defunct parties. }
\end{center}
\end{figure}

\begin{figure}
\begin{center}
\includegraphics[width=0.75\textwidth]{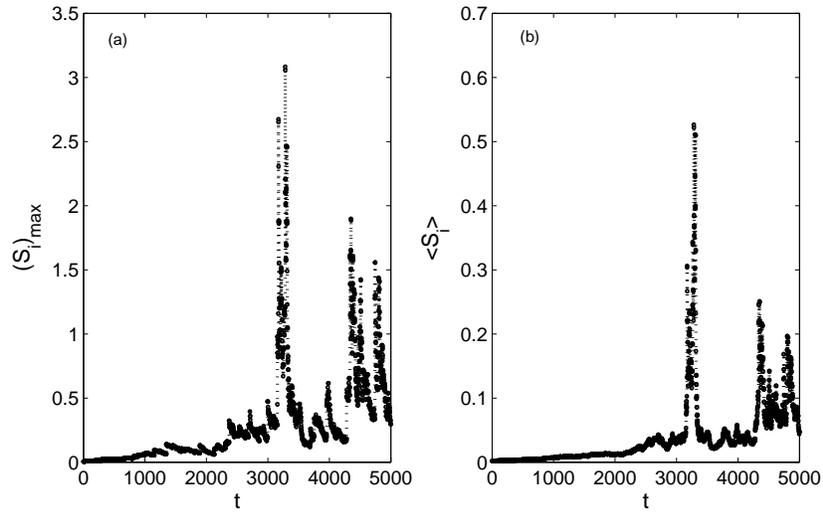}
\caption{Time evolution of (a) the maximum solidarity $(S_{i})_{max}$ and (b) the average solidarity $<S_{i}>$. The first peak of $(S_{i})_{max}\approx 3$ is at about $t=3200$. }
\end{center}
\end{figure}

\begin{figure}
\begin{center}
\includegraphics[width=0.75\textwidth]{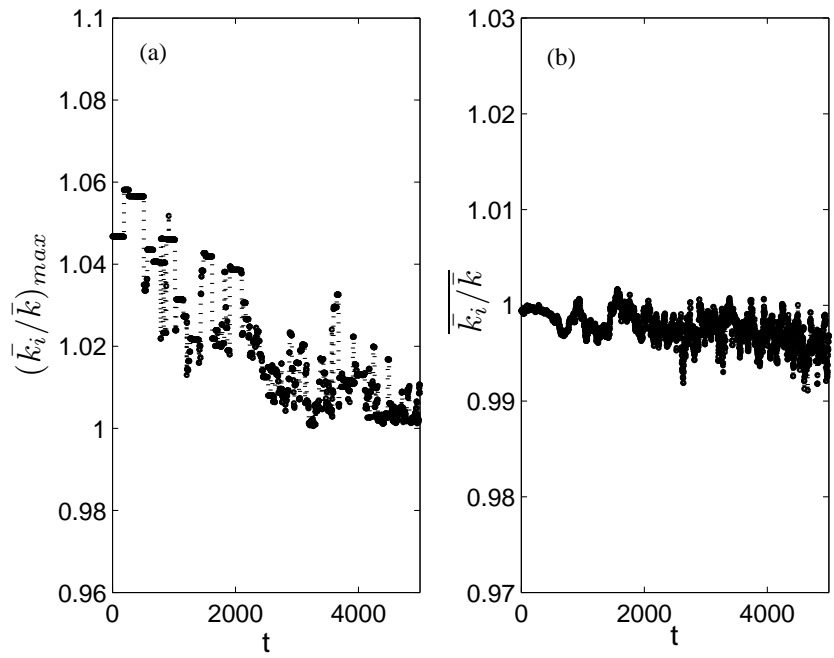}
\caption{Time evolution of (a) the maximum average relative degree and (b) the mean of the average relative degree. Both of them fluctuate around 1 at long time.}
\end{center}
\end{figure}

\begin{figure}
\begin{center}
\includegraphics[width=0.75\textwidth]{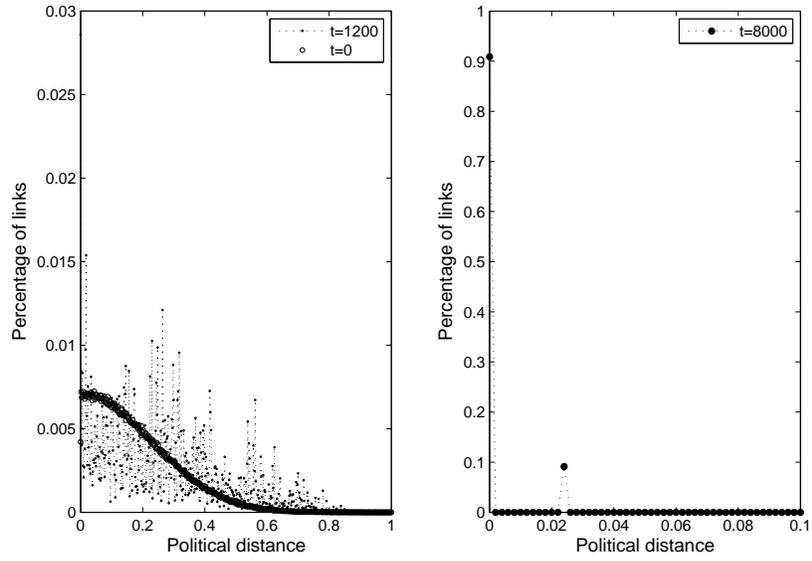}
\caption{Political distance distribution of links in three time steps. The distribution changes from initial semi-Gaussian distribution ($t=0$) to the final semi-delta type distribution around zero distance ($t=8000$).}
\end{center}
\end{figure}

\begin{figure}
\begin{center}
\includegraphics[width=0.75\textwidth]{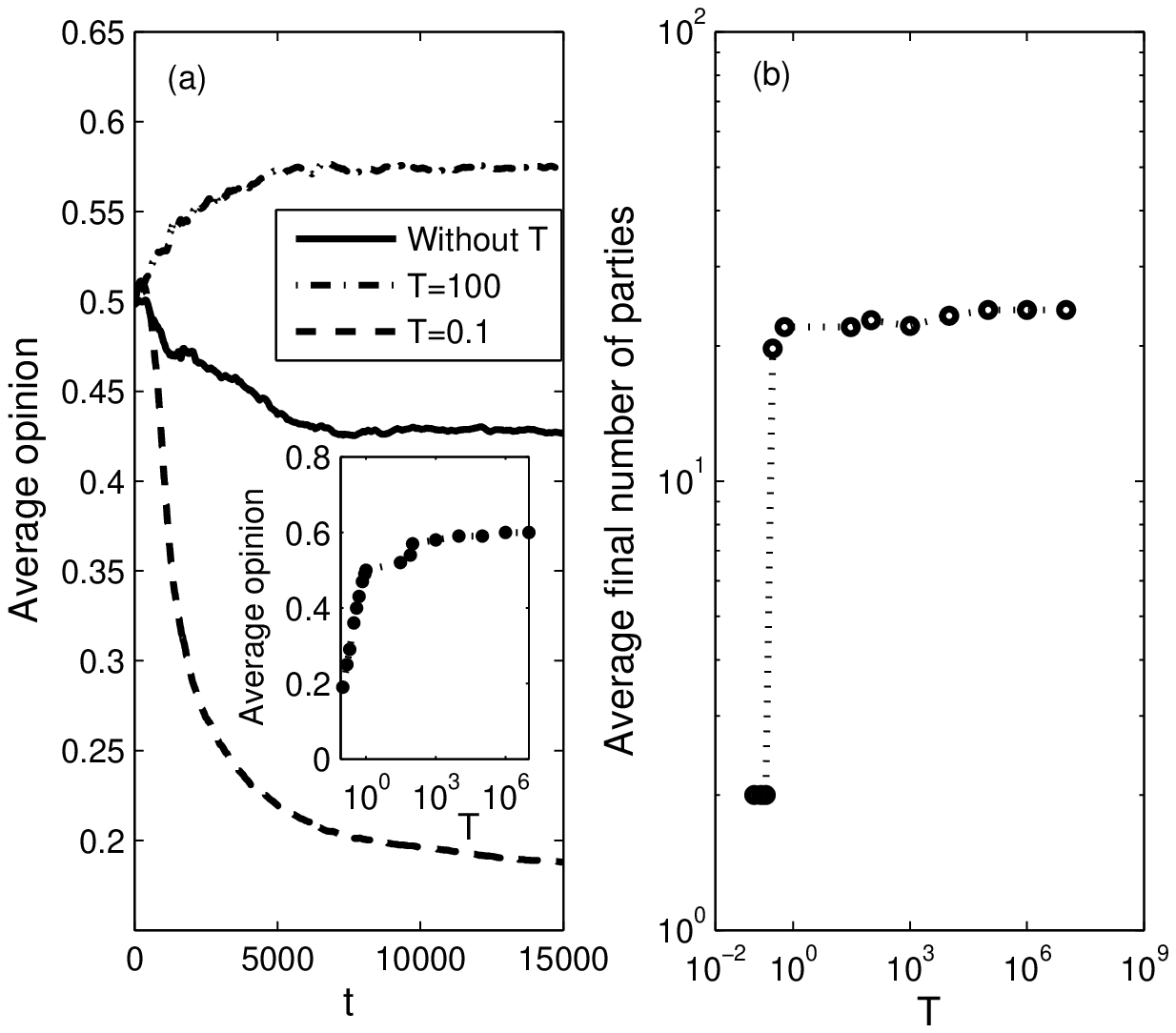}
\caption{(a) Time evolution of the average opinion $\bar{x}$ at two different temperatures in comparison with the evolution without thermostat. Note that for $T=0.1$, The stable $\bar{x}$ is about 0.17, so that $T<\bar{x}$. The inset shows the temperature dependence of the stable average opinion. (b) Variation of the average number of remaining parties v.s. temperature. For $T$ larger than 0.2, $<n_t>$ is about 26 for $\alpha=0.1$, $N=5000$, $n=500$, and $TR=2$. $<n_t>$ is 2 when $T$ is smaller than 0.2.}
\end{center}
\end{figure}


\begin{thebibliography}{99}
\bibitem{Comment1}
This average opinion is an arithmetic one given by $\left\langle x\right\rangle=\frac{\sum_{i=1}^{n_t}x_i}{n_t}$ where $n_t$ is the number of parties at time $t$. We also tried the average weighted with the numbers of agents of the parties. But the result is similar.
\bibitem{Comment2}
A mean position weighted by the agent numbers of each party may be more judicious but this will not change the main evolutionary characters of the system.
\end{thebibliography}
\end{document}